\documentclass[aps,amsmath,amssymb, reprint]{revtex4-1}
\usepackage{xcolor}
\usepackage{graphicx}
\usepackage{dcolumn}
\usepackage{bm}
\usepackage{soul}
\usepackage[utf8]{inputenc}
\usepackage[T1]{fontenc}
\usepackage{mathptmx}
\usepackage{float}
\usepackage{lipsum}
\usepackage{caption}
\usepackage{subcaption}
\usepackage{amsmath}
\usepackage[normalem]{ulem}

\begin{document}

\title{Rheology of {\it Pseudomonas fluorescens} biofilms: from experiments to DPD mesoscopic modelling }

\author{Jos\'e Mart\'in-Roca}
\affiliation{ Dep. Est. de la Materia, F\'isica T\'ermica y Electr\'onica, Universidad Complutense de Madrid, 28040 Madrid, Spain}

\author{Valentino Bianco}
\affiliation{ Departamento de Quimica Fisica, Facultad de Ciencias Quimicas, Universidad Complutense de Madrid, 28040 Madrid, Spain}

  \author{Francisco Alarcón}
    \affiliation{Departamento de Estructura de la Materia, F\'isica T\'ermica y Electr\'onica, Facultad de Ciencias F\'isicas, Universidad Complutense de Madrid, 28040, Madrid, Spain}
    \affiliation{Departamento de Ingeniería Física, División de Ciencias e Ingenierías, Universidad de Guanajuato, Loma del Bosque 103, 37150 León, Mexico.}
    
\author{Ajay K. Monnappa}
\affiliation{ Departamento de Quimica Fisica, Facultad de Ciencias Quimicas, Universidad Complutense de Madrid, 28040 Madrid, Spain}
\author{Paolo Natale}
\affiliation{ Departamento de Quimica Fisica, Facultad de Ciencias Quimicas, Universidad Complutense de Madrid, 28040 Madrid, Spain}
\author{Francisco Monroy}
\affiliation{ Departamento de Quimica Fisica, Facultad de Ciencias Quimicas, Universidad Complutense de Madrid, 28040 Madrid, Spain}

\author{Belen Orgaz}
\affiliation{ Sección Departamental de Farmacia Galénica y Tecnología Alimentaria, Facultad de Veterinaria, Universidad Complutense de Madrid, Madrid, Spain}

\author{Ivan L\'opez-Montero}
\affiliation{ Departamento de Quimica Fisica, Facultad de Ciencias Quimicas, Universidad Complutense de Madrid, 28040 Madrid, Spain}
\affiliation{Instituto de Investigación Biomédica Hospital 12 de Octubre (imas12), Madrid, Spain}

 \author{Chantal Valeriani}
\affiliation{ 
Dep. de Est. de la Materia, F\'isica T\'ermica y Electr\'onica, Universidad Complutense de Madrid, 28040 Madrid, Spain
}
\affiliation{ 
GISC - Grupo Interdisciplinar de Sistemas Complejos 28040 Madrid, Spain
}

\begin{abstract}
The presence of bacterial biofilms in clinical and industrial settings is a major issue worldwide. A biofilm is a viscoelastic matrix, composed of bacteria producing a network of Extracellular Polymeric Substances (EPS) to which bacteria crosslink. 
Modelling of complex biofilms is relevant to provide accurate descriptions and predictions that include parameters as hydrodynamics, dynamics of the bacterial population and solute mass transport. However, up-to-date numerical modelling, even at a coarse-grained level, is not satisfactorily. In this work, we present a numerical coarse-grain model of a bacterial biofilm, consisting of bacteria immersed in an aqueous matrix of a polymer network, that allows to study rheological properties of a biofilm. 
We study its viscoelastic modulus, 
varying topology and composition 
(such as the number of crosslinks between EPS polymers, the number of bacteria and the amount of solvent), and  compare the numerical results with experimental rheological data of {\it Pseudomonas fluorescens} biofilms grown under static and shaking conditions, as previously described by Jara et al, Frontiers in Microbiology (2021).

\end{abstract}

\maketitle

\section{Introduction}

Biofilms are synergistic colonies of one or more  bacterial species growing on a solid surface most often in contact with an aqeuous liquid.
In contrast to liquid-floating planktonic bacterial cells, bacteria embedded in solid-supported biofilms do not have unlimited access to nutrients. 
The presence of biofilms does not only provide the physical protection of the embedded  cells against chemical or mechanical challenges\cite{peterson,fleming2010}, but it also facilitates the evolution of phenotypes that eventually lead to the resistance to the treatment of antimicrobial agents \cite{Donlan,Hoiby2010}. Antimicrobial resistance of bacteria dwelled in biofilms compromises the process of bacterial removal giving rise to a more persistent bacterial infection. 
Reason why biofilm formation is 
a major health issue \cite{Khatoon2018, Milchev2013} and industrial challenge worldwide \cite{Duan2008,Mansour2016,Schultz2000}. 
Bacterial infections in plants and animals  lead to food contamination and poisoning\cite{Zhao} 
. 
\noindent
A biofilm is a complex and heterogeneous viscoelastic material \cite{wilking2011} produced by bacterial cells. 
In more structural terms, a biofilm  consists of 
a network of extracellular polymeric substances (EPS) composed of polysaccharides, proteins,  extracellular DNA, and cells, everything surrounded by  an aqueous environment. 
EPS  serves as a scaffold for matrix cross-linked bacterial cells embedded within the biofilm. 
The amount  of bacteria with respect to  EPS  vary from one tenth to one fourth of the total biofilm mass\cite{garrett2008,Sveinbjornsson2012}, depending on the bacterial species.
\noindent
Mathematical modelling of bacterial biofilms started more than 30 years ago \cite{klapper2010,Wang2010}.  Biofilms have been described via phase-field (continuum) models, in which one phase represented the biofilm (EPS and bacteria) and the other phase the surrounding  matrix containing the nutrient substrates \cite{zang2008phase}. 
 \noindent
The tension and compression of a biofilm have been computed using finite elements simulations to obtain the Young's modulus of an artificial biofilm composed of bacteria  embedded in an agarose hydrogel \cite{kandemir}.

 \noindent
A different approach consisted in implementing an hybrid discrete-continuum models to incorporate the flow of substrate, bacterial growth or biomass spreading over the rough biofilm surfaces \cite{Picioreanu1999,Picioreanu2004}. More recently, mesoscale numerical models such as the immersed boundary-based method (IBM) or the dissipative particle dynamics (DPD) have been proposed to study biofilm maturation \cite{Horn2014,Eberl2008,Liu2018,Xu2011}, establishing a connection between the mechanical stress and strain of the biofilm \cite{Bol2012,Hammo2014,Stotsky2016}. 
\noindent
On the one side, to model a complex system via the IBM model  \cite{peskin1,peskin2} several assumptions need to be taken into account, such as 1) not considering  biofilm growth but steadiness during the simulation time scale \cite{IBM0}, 2) mimicking  biofilm elasticity via linear springs that connect individual bacterial cells \cite{IBM} and 3) capturing the biofilm viscosity via constitutive equations for stress.
As a matter of phenomenological evidence, macroscopic rheological experiments were considered on {\it Staphylococcus epidermidis}
biofilms grown on a plate rheometer \cite{pavlovsky2013situ}. They were compared qualitatively to numerical results obtained via the microscale spatial IBM model  \cite{IBM2}. 
\noindent
On the other side, as refers to dynamic methods \cite{barai2016modeling}, DPD conserve mass and momentum under steady state conditions.
Differently from grid-based computational methods  that do not consider any mechanical constraints, DPD can be easily implemented when considering objects  with complex geometries \cite{xu2011dissipative}. 
Like Molecular Dynamics (MD) simulations \cite{mukhi2021identifying}
, DPD captures the time evolution of a many-body system with friction explicitly considered. Unlike MD, DPD allows to model physical phenomena occurring at longer time scales and length scale \cite{raos2011}. 
DPD was thus implemented 
 to simulate  biofilm formation on post-coated surfaces \cite{Balazs}, and the growth of two-dimensional biofilm in flow \cite{Zhijie}.
 \noindent
 Recently, we have presented a combined numerical and experimental study on  how hydrodynamic stress affect  structural features of {\it Pseudomonas fluorescens} biofilms \cite{jara2021}. By means of rheological measurements on a biofilm grown in the presence and in the absence of hydrodynamic stress, we demonstrated that
 bacteria are capable of adapting to hostile deformations by tailoring 
  the  structure of the matrix and its viscoelastic properties \cite{jara2021}.
    The results were supported by numerical results of the rheological properties  of a mimetic biofilm simulated as a first approximation via  a DPD model. 
    Thanks to computer simulations, we learned that biofilms grown under stress were mechanically more stable due to an increase of the number of crosslinks between EPS polymers.


\noindent
In this work 
we probe the rheological features of a {\it P. fluorescens} biofilms 
varying the frequency of the externally imposed shear flow. 
 To  numerically study the biofilm viscoelastic modulus, 
we use a coarse-grain  DPD model. To better characterised the biofilm, we 
vary  topology and composition parameters 
such as the number of matrix crosslinks between EPS polymers, the number of bacteria embedded  and the amount of solvent. 
The numerical results are compared    to experimental rheological data of {\it P. fluorescens} biofilms grown under static and shaking conditions.


\section{Numerical and experimental details}
In what follows, we present the mesoscopic model  used to simulate mature biofilms, and the experimental details of the viscoelastic measurements of a {\it P. fluorescens} biofilm. 
\subsection{Mesoscopic model for an {\it in silico} biofilm}




{\color{black}
The simulated biofilm consists of three different components: bacteria, polymers and solvent.} 
The biofilm is prepared in an  initial configuration consisting of a colony of monodisperse bacteria immersed in a periodic box containing solvent and randomly distributed polymers.

\noindent
\subsubsection{Bacteria}
Differently from the work described by Raos \textit{et al}.\cite{GuidoRaos2006}, where   dispersed filler particles aggregated into a single bacterium, we use realistic bacteria consisting of already clustered particles. 
We model the bacterium as a bacillus (rod-shaped) composed by 440 particles (of type $b$), each of diameter $\sigma_b$, held together via harmonic interactions.
\begin{equation}
 U_{b}^{\rm bond}=\frac{K_b}{2}(r_{i,j} - \sigma_b)^2 \qquad,
\end{equation}
being $r_{i,j}$ the distance between the bonded beads $i$ and $j$, $\sigma_b$ the equilibrium distance, $K_b$ the harmonic coupling constant. The central part of a bacterium is shaped as an empty cylinder of length $l_b/\sigma_b = 6$, while the extremes of the bacteria are shaped as spherical caps of radius $R_{ext}/\sigma_b = 4$. Both the central and extreme parts have an external radius $R_{ext}/\sigma_b = 4$ and internal radius $R_{ext}/\sigma_b =2$, i.e. the bacterium membrane is formed by three layers of particles.

\noindent
\subsubsection{Matrix: polymers and solvent}
The bacteria are immersed in a bulk matrix made of polymers (EPS) and solvent. The solvent is represented by $N_{\rm solv}$ particles of type $s$, whereas the EPS is represented by linear chains of $l$ beads of type $p$ that are bound via harmonic potential: 
\begin{equation}
 U_{p}^{\rm bond}=\frac{K_p}{2}(r_{i,j} - \sigma_p)^2 \qquad,
\end{equation}
where $r_{i,j}$ is the distance between two consecutive beads $i$ and $j$, $\sigma_p$ the equilibrium distance, and $K_p$ the coupling constant.

\noindent
\subsubsection{Cross-links and non-bonded interactions}
The crosslink (CL) interactions, occurring between two beads of
type $p$ (polymer-polymer crosslink), or between beads of type $b$ and
$p$ (bacterium-polymer crosslink), are described via harmonic potential:
\begin{equation}
 U^{\rm CL}=\frac{K}{2}(r_{i,j} - \sigma_{CL})^2 \qquad.
\end{equation}

\noindent
All non-bonded interactions, namely polymer-polymer (pp), polymer-bacterium (pb), polymer-solvent (ps), bacterium-bacterium (bb), bacterium-solvent (bs), and solvent-solvent (ss) interactions, are modelled according to the DPD force field. 
In particular, the force between two particles $i$ and $j$, respectively of type $\alpha$ and $\beta$, is expressed as the sum of a conservative force $\vec{F}^C_{i,j}$, a dissipative force $ \vec{F}^D_{i,j}$ and a random force $\vec{F}^R_{i,j}$ given by 
\begin{eqnarray}
\label{eq:dpdcons}
 \vec{F}^C_{i,j} = B_{\alpha,\beta}w(r)\hat{r}_{i,j} \qquad r<r_c 
 \end{eqnarray}
  where $r_c$ is the cut-off distance, beyond which all the terms vanish;  $\hat{r}_{i,j}$ is the unit vector along the direction of the vector $\vec{r}_i -\vec{r}_j$, indicating the positions of  particles $i$ and $j$, respectively; $B_{\alpha,\beta}$ is the amplitude of the conservative force between  particles of type $\alpha$ and $\beta$; 
 \begin{eqnarray}
 \vec{F}^D_{i,j} = -\gamma w^2(r)\left(\hat{r}_{i,j}\cdotp \hat{v}_{i,j}\right)\hat{r}_{i,j} 
 \end{eqnarray}
 where   $\vec{v}_{i,j}\equiv\vec{v}_i-\vec{v}_j$ is the vector difference between the  velocity $\vec{v}_i$ of  particles $i$, and the  velocity $\vec{v}_j$ of  particle $j$; $\gamma$ the friction coefficient; 
  \begin{eqnarray}
 \vec{F}^R_{i,j} = w(r) \left(\dfrac{2k_BT\gamma}{dt}\right)^{1/2} \Theta \, \, \, \hat{r}_{i,j} \\
 w(r) = 1-r/r_c
\end{eqnarray}
Here,  $w(r)$ is a weighting factor varying from 0 to 1;  
$\Theta$ is a Gaussian random number with zero mean and unit variance; $dt$ the integration time;  $k_B$ the Boltzman constant;  $T$ the absolute temperature. 
The sum of the three DPD forces allows to capturing  the long-range correlations induced by the hydrodynamic interactions, also accounting for thermal fluctuations.


\subsection{Numerical details}
We simulated the biofilm via the LAMMPS open source code\cite{lammps}.
The system evolves according to a velocity Verlet algorithm, keeping fixed the volume of the simulation box, set  to $32\times32\times32$ with periodic boundary conditions.
We simulate the biofilm  changing the number of bacteria $N_b$ from 90 to 184, the number of polymers $N_p$ (from 80 to 240), the polymer lengths $l$ (from 50 to 200) or the number of solvent particles $N_s$ (from 35000 to 75000). All parameters $N_p$, $N_s$ and $N_b$ are chosen to guarantee the DPD condition for the density\cite{dpd,groot1997dissipative}:  $(N_p + N_b + N_s ) / ( 32 \times 32 \times 32) > 3$. Using internal units, we set $r_c=1$ for all DPD interactions, $\sigma_b=\sigma_p=0.5$, $\gamma=4.5$, $K_b=K_p=30$. The time step is set to $dt=0.05$, although we tested our results also against $dt=0.005$. 
Moreover, to guarantee that bacteria and polymers are properly hydrated, we choose the amplitudes of the solvent interactions in eq.\ref{eq:dpdcons}
smaller than the remaining ones. In particular, we choose $B_{s,s}=B_{s,b}=B_{s,p}=25$ and $B_{b,b}=B_{p,p}=B_{p,b}=30$. 
\noindent
To convert those dimensionless units in real world units, we consider $k_BT=4.11\times 10^{-21}$ J as the characteristic energy scale in our simulation. Also, we set the system length scale to the longest dimension of one \textit{P. fluorescens} bacterium, which can be considered as approximately  $\sim 1.5 \mu$m.  
In our simulation the bacterium is 15 beads long, which results in a unit length corresponding to $l_u \sim 200$ nm (hence $\sigma_b=\sigma_p \sim 100$ nm). 
Moreover, since the mass of a single bacterium is $\sim 10^{-15}$ Kg, the mass of a single particle is set to $m_u \sim 2.3\times 10^{-18}$ Kg. The  time scale is derived accordingly as $\tau_{\rm intrinsic} = \sqrt{m_ul_u^2/k_BT}\sim 5\times10^{-6}$s. 
At a time step of dt=0.01, a period of T=200 corresponds to 2000 time steps that is within  the total lifespan of the simulation runs. 


\subsection{{\bf Setting up the initial configuration}}
We first choose the parameter space to explore. 
 By fixing $N_b$, $N_p$, $l$ and $N_s$, we equilibrate the system in the absence of polymer cross-links (CL) for $\sim 8\times 10^6$ time steps and perform  simulations  changing the parameters within the following ranges: $N_b=$ 110, 120, 130, 140, 150, 160, 175, 184, 190, 195; $N_p=$ 50, 70, 80, 90, 110, 140, 170; $N_s=$ 20000, 30000, 35000, 40000, 50000, number of CL=100, 300, 500, 700, 900, 1100, 1300, 1500, 1700, 2000, 2300, 2400, 2800, 3000, 3200, 3300, 3600, 3800, 4000, 4300, 4400, 4800, 5300, 6000, 7000.

\begin{figure}[h!]
\centering
 \includegraphics[width=0.3\textwidth]{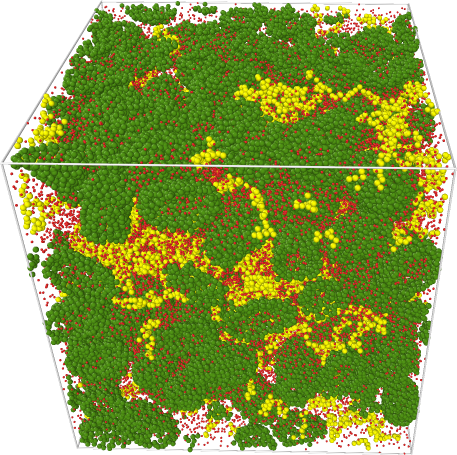}
 \caption{{\color{black}Snapshots of the simulation box containing 184 bacteria (in green), 80 polymers (in yellow) and 35000 solvent particles (in red). 
 The size of a single particle has been chosen for the sake of visualization and does not reflect the interaction radius.}}
 \label{fig:model}
\end{figure}

\noindent
Having prepared the initial configuration, we need to locate the crosslinks within the polymer network. 
To create the network of cross-linked polymers and bacteria, we randomly place $N_{\rm CL}$ new harmonic bonds 1) between $b$ and $p$ types of particles, i.e. a bacterium-polymer CL and 2) between two $p$ particles (belonging or not to the same polymer chain), i.e. a polymer-polymer CL.
To avoid crosslinks agglomerating between nearest neighbours, leading to a clump of $p$ particles of the same polymer chain, we 
forbid CL between the ten closest neighbors along the same chain.
Therefore,  we fix the  intermolecular distance of the polymer loops within one polymer (\textit{intra}) or between different polymers (\textit{inter}) to be at least 10 $\sigma$ apart. 
To prevent the formation of too many cross-links per particle which would result in a globule-like clusters of beads, we assume that 
any particle of type $b$ can form at most one CL, while any $p$ particle can form at most two CL. 
 \noindent
 For any choice of $N_b$, $N_p$, we create an initial configuration and  form a number of CL ranging from 2300 up to 5400. 
After the formation of the CLs, a second equilibration run of $\sim 10^6$ time steps is performed to relax the polymer-bacteria network. 
\noindent
A characteristic snapshot of the biofilm  is shown in Fig. \ref{fig:model}. 

\noindent
\subsection{Simulated rheology: numerical details}
To study the rheological properties of the biofilm, we compute the shear modulus by 
monitoring the shear stress response $\sigma(t)$ under a sinusoidal deformation. 
We apply an external oscillatory shear deformation with frequency $\nu$ along th $X$--$Y$ plane by changing the box size $L_z$\cite{raos2011,jara2021} according to
\begin{equation}
L_z (t)= L_0 + A\sin(2\pi \nu t) \qquad,
\label{eq:shear_zsize}
\end{equation}
being $L_0=32$ the initial box size and $A$ the amplitude of the oscillation. Under shearing conditions, the particle velocities are remapped every time  they cross periodic boundaries. Within our simulations, we have explored the following values for the amplitude $A$ :A= 2.5, 3.2, 4.8, 8, 12, 16, 20, 24, 28, 32, 36; and the period $T$: T$\equiv\frac{2\pi}{\nu}$= 20, 50, 100, 200, 500, 1200, 2000, 2500, 5000, 10000, 20000, 100000, 200000.

\noindent
According to Raos \textit{et al}.  \cite{GuidoRaos2006}, the resulting stress has been calculated by fitting the $xy$ component of the shear stress with: 
\begin{equation}
 \sigma_{xy}(t)=\sigma'\cdot \sin(2\pi \nu t) + \sigma''\cdot \cos(2\pi \nu t) 
 \label{eq:sigma_xy}
\end{equation} 
\noindent
and the amplitude values of $\sigma'$ and $\sigma''$ were used to compute the the in-phase and out-of-phase components of the complex shear modulus
\begin{equation}
G' = \frac{\sigma'}{(A/L_0)} \qquad G'' = \frac{\sigma''}{(A/L_0)}
\end{equation}
being $G'$ the storage modulus and $G''$ the loss modulus, respectively.
 Rheological properties of solid biomaterials 
 are characterized by a finite storage modulus $(G^{'}>0$). Whereas fluid viscous materials are characterized
by near zero storage modulus $(G^{'} \sim 0$) and a large loss modulus $(G^{''}>0$)\cite{donlan2001biofilms,alonci2018injectable}.

\noindent
From a phenomenological point of view, the computed stress-strain curves are mainly characterized by two regimes 
a) The stress-strain curves are linear at low strain, where the stress is proportional to the the strain. 
b) Beyond a yield strain, where the stress-strain curves deviate  from the linear regime. 
Both regimes crossover at the yield point, which fixes the yield strain in which the system softens becoming effectively plastic.
 
\noindent
For the considered frequency ranging from  a value of $3\times10^{-5}$ to a value of $0.3$ (in internal units),
we run the simulations completing at least 3 box-shearing oscillations. We consider simulations amplitudes $A$ (ranging from 4 to 36), corresponding to a deformation of up to the 112\% of the box size. 
However, we will mainly focus on data obtained for $A=24$ (75\%), since:
the stress-strain curve is linear until around a deformation of 50\% and starts being non linear for a larger deformation; 
when varying all parameters ($N_b$, $N_p$, ...) while the behaviour of $G^{'}$ and $G^{''}$ is not strongly affected by the choice of A, their response is stronger and the noise lower for $A=75\%$.
Thus, we have considered a period $T$ ranging from $2 \times 10$ to $2 \times 10^5$. 


\noindent
\subsection{{\bf Experimental details: Microorganisms and growth conditions}} 
The strain \textit{P. fluorescens} B52, originally isolated from raw milk \cite{Richardson1978}, was used as model microorganism. 
Overnight precultures and cultures were incubated at 20$^\circ$C under continuous orbital shaking (80 rpm) in tubes containing 10 ml Trypticase Soy Broth (TSB, Oxoid). Cells were recovered by centrifugation at 4,000 x \textit{ g} for 10 min 
(Rotor SA-600; Sorvall RC-5B-Refrigerated Superspeed Centrifuge, DuPont Instruments)
and washed twice with sterile medium. Cellular suspensions at OD$_{600}$ were first adjusted to 0.12 (equivalent to 10$^8$ cfu/ml) 
and then diluted  to start the experiments at  10$^4$ cfu/ml. 

\noindent
\subsection{Experimental details: Biofilm development}
Biofilms were developed using borosilicate glass surfaces ($20\times 20$ cm) as adhesion substrate. Five glass plates were held vertically into the sections of a tempered glass separating chamber provided with a lid. The whole system was heat-sterilized as a unit before aseptically introducing 2 mL of the inoculated culture medium. To check the effect of hydrodynamic stress on biofilm mechanical properties, incubation was carried out for 96 h at 20$^\circ$C both in an orbital shaker at 80 rpm (shaken sample) and statically (static sample). For biofilm recovery, plates were aseptically withdrawn, rinsed with sterile saline in order to eliminate weakly attached cells and then scraped to remove the attached biomass (cells + matrix) from both faces of the plates. For rheology measurements, biofilm material was casted every 24 h to be directly poured onto the rheometer plate. Experiments were run in triplicate.

\noindent
\subsection{Experimental rheology} 
As previously described \cite{jara2021}, 
the viscoelastic response was determined in a hybrid rheometer under oscillatory shear stress-control (Discovery HR-2, TA instruments). We used a cone-plate geometry  (40 mm diameter) and a Peltier element to control the temperature through an external an external water thermostat.
Measurements were performed at a 1 mm gap between the Peltier element
and the plate tool (TA instruments), where a sinusoidal strain 
$\gamma$ of amplitude $\gamma_0$ was performed at a frequency  $\omega$, as  $\gamma (t) = \gamma_0 sin (\omega t)$. The shear stress exerted on the biofilm was monitored as $\sigma(t) = G^{*} \gamma (t)$, where 
 $G^{*}$ is the viscoelastic modulus $G^{*} = G^{'} + i G^{''}$, where $G^{'}$ is the storage modulus accounting for shear rigidity, $G^{''}$ is
the loss modulus  accounting for viscous friction. These properties were considered under the same definitions as in Eq. (10). 

\section{Results}

\subsection{Matching the parameters of the biofilm model}
\begin{figure}[h!]
 \includegraphics[width=0.4 \textwidth]{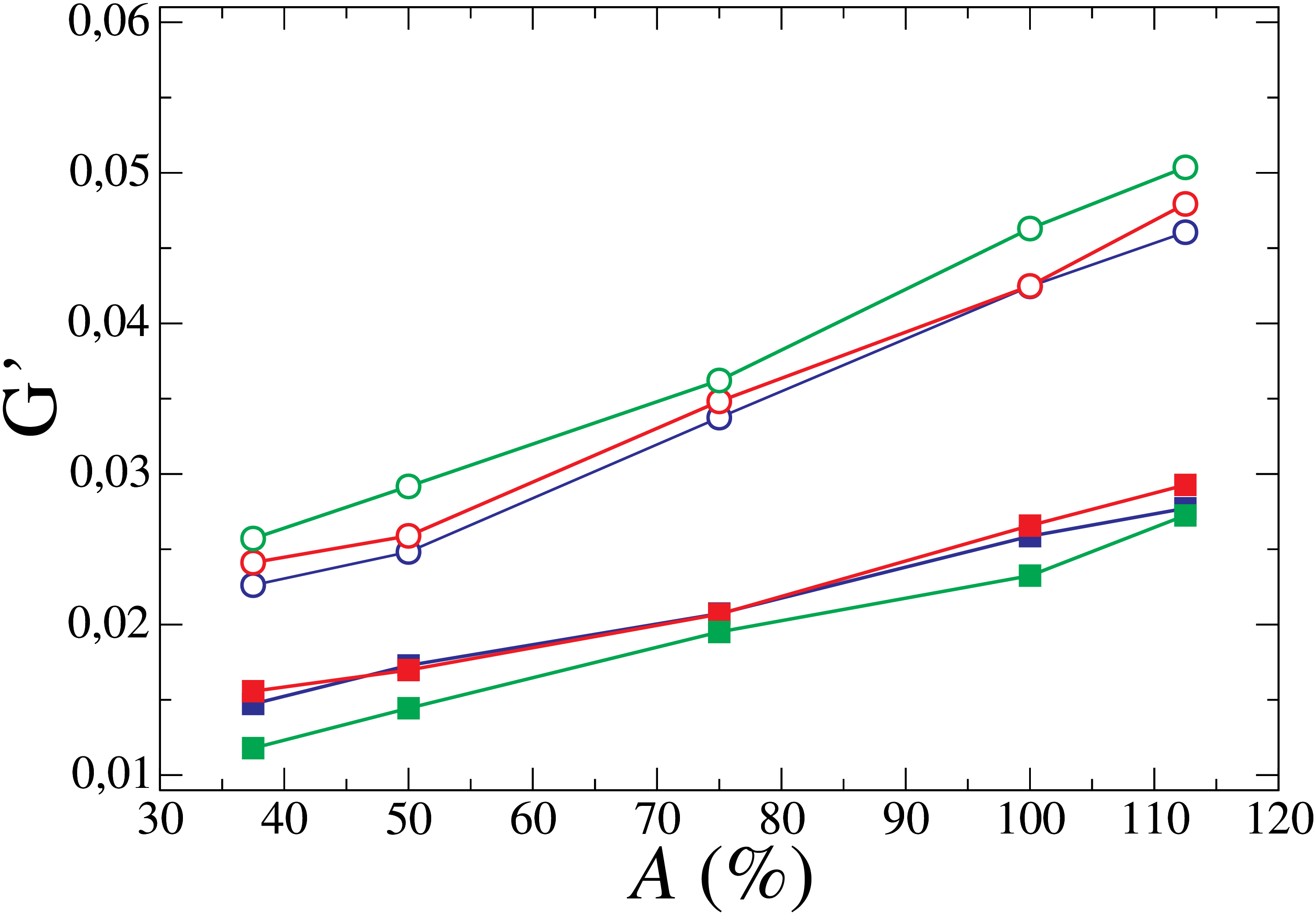} 
 \caption{Storage Modulus as a function of the amplitude for different cross-link topologies in the presence (filled squares) and absence (empty circles) of polymer loops. The color code indicates different numbers of polymer-polymer (pp), polymer-bacteria (pb) cross-links: $N_{CL,pp}$ =200/ $N_{CL,pb}$=1400 (blue); $N_{CL,pp}$ =400/ $N_{CL,pb}$=1200 (red) and $N_{CL,pp}$ =600/ $N_{CL,pb}$=1000 (green). 
 All  data refer to $N_b=150$, $N_p=80$, $N_s=45000$ and $T=2\times10^5$.
 } 
 \label{stress_tropology} 
 \end{figure}

 \noindent
 To start with, we study  the stress response of the biofilm  containing cross-linked polymer loops
regardless of their intermolecular distances along the polymer chain. 
 We analyze how parameters affect the in-phase ($G^{'}$) 
stress response function. In Fig. \ref{stress_tropology} we represent the storage modulus $G^{'}$ as a function of the shear amplitude, keeping fixed $N_b=150$, $N_p=80$, $T=2 \times 10^{5}$, $l=100$ and the total number of $CL=1600$ cross-links ($N_{CL,pp} + N_{CL,pb}$).

  \begin{figure*}
 \makebox[\textwidth][c]{\includegraphics[width=0.87\textwidth]{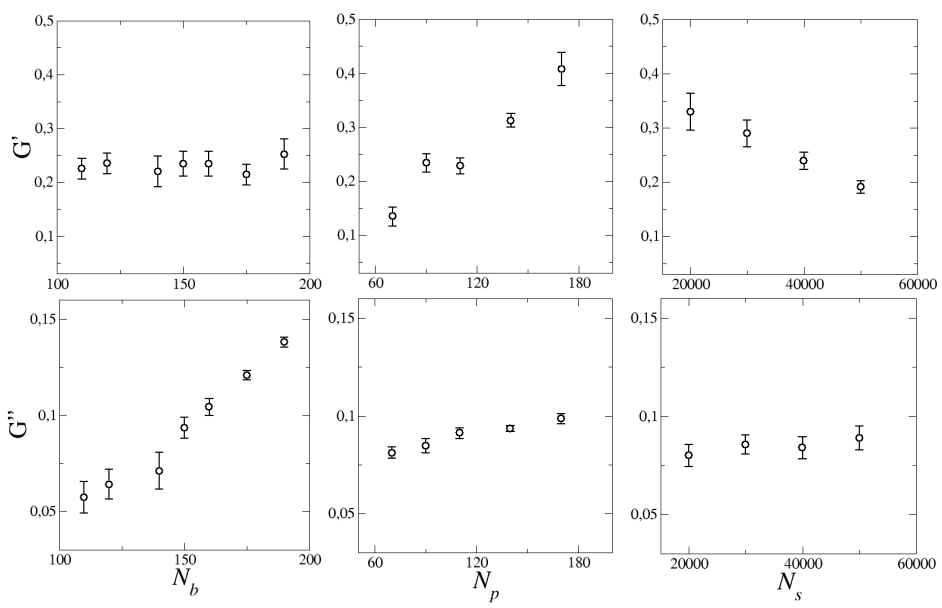}}%
 \caption{Stress components $G'$ and $G''$ for $A=75\%$, $CL=2300$ and $T=200$ as function of (a,d) the number of bacteria $N_b$; (b,e) the number of polymers $N_p$ and solvent (c,f).}
 \label{fig:Loss_Storage_Modulus}
 \end{figure*}
 
 \noindent
 We consider two systems with different cross-link topologies:  one with polymer cross-links (filled squares) and one without polymer cross-links  (empty circles). The presence of polymer loops between $p$ particles belonging to the same polymer chain affects the storage modulus. Since $G'$  of the cross-linked biofilm (filled squares, in Fig.\ref{stress_tropology}) 
is lower than the one of  the cross-linked biofilm that does not include polymer loops (Fig.\ref{stress_tropology}, empty symbols).

 This finding indicates the importance to include polymer loops with a biofilm model. As expected, independently of the presence (or absence) of polymer loops, the storage modulus increases with the strain-amplitude. In the presence of polymer loops (for the chosen fixed parameters), we observe that the shear response is not affected by the ratio between polymer-polymer ($pp$) versus polymer-bacteria ($pb$) cross-links. On the other hand, when polymer loops are not present within the same polymer chain, the system response increases (empty symbols in Fig. \ref{stress_tropology}). We think that this increase in system response is due to the presence of a larger number of interactions (connections) between bacteria and matrix polymers. This  leads to a a denser and stiffer biofilm and as a consequence  generates a larger response function.



\noindent
\subsection{The shear stress dependence on $N_b$, $N_p$ and $N_s$}
\noindent
To check whether the shear stress depends on the number of bacterial cells ($N_b$), in Fig. \ref{fig:Loss_Storage_Modulus}, panel a)we study the $G^{'}$ (in-phase) and $G^{''}$ (out-of-phase) response as a function of $N_b$ (fixing all other parameters). 
We use an intermediate number of cross-links ($CL=2300$), and intermediate values $N_p=80$, $l_p=100$, $N_s=35000$ within the explored range. 
Moreover, we apply a deformation amplitude of $A=75$ $\%$ and a period of the deformation of $T=200$. 
Fixing the different oscillation's periods to $T=2 \times 10^2$, we  observe a relevant response of $G^{''}$. Whereas at lower frequencies we  have found an amplitude ($A$) depending limiting value of $G^{''}$ (as we will also observe later in  Fig \ref{fig:fig_exp}).

\noindent
Our results show that the in-phase response (accounting for the storage, or rigidity modulus $G^{'}$) remains constant through of the $N_b$ range (Fig. \ref{fig:Loss_Storage_Modulus}, panel a). This result is in agreement with previously reported data \cite{wilking2011}. 
Considering the bacterial biofilm as a polymer-colloid viscoelastic medium \cite{wilking2011}, $G^{'}$ should increase with the bacterial density until reaching a solid-like plateau. Beyond this point, the system behaves as a viscous (liquid-like) fluid. Our findings point out that the probed $N_b$-range falls into the elastic response regime, where $G^{'}$ is not affected by $N_b$.

\noindent
The loss modulus $G^{''}$ increases linearly with $N_b$ (Fig.\ref{fig:Loss_Storage_Modulus}, panel d), reflecting the high friction imposed on the system by dragging the large bacteria while  increasing their density.
The viscous loss almost reaches the values of the storage modulus for the largest number of bacteria considered. 
However, since $G^{''} $ is always smaller than $G^{'} $, the biofilm system behaves as a pasty material, with a relative high structural rigidity $G^{'} > G^{''} $ instead of a viscous fluid where $G^{'} = 0$ and $G^{''} > 0$. Thus our model shows a rheological behavior of a quite resilient material with a relatively high mechanical compliance.


\noindent
To visualize the effect of an increasing number of entangled polymers on the shear response, we consider $N_b=150$, $l_p=100$, $N_s=35000$, $CL=2300$ with an amplitude of $A=75$ $\%$ and period of the deformation of $T=200$ as boundary conditions for our simulations (Fig.\ref{fig:Loss_Storage_Modulus}, panel b and e). 
When we increase the number of polymers  and maintain a constant bacterial density, we observe that it is easier for the biofilm to generate a solid-like network with an increased rigidity (Fig.\ref{fig:Loss_Storage_Modulus}b). This suggests that bacteria colonize a preexisting polymer template and modify it according to their necessities.
The out-phase stress response  increases with increasing number of polymers (Fig.\ref{fig:Loss_Storage_Modulus} e) or solvent (Fig.\ref{fig:Loss_Storage_Modulus} f) present in the biofilm. Since this the increase of $G^{''}$ is quite small when compared to the response caused by the presence of bacteria (Fig.\ref{fig:Loss_Storage_Modulus} d), we consider $G^{''}$ for polymer and solvent to be constant.

\noindent
When studying  the mechanical response of the system for different hydration levels ($CL=2300$, $N_b=184$, $N_p=80$, $l_p=100$ with an amplitude of $A=75$ \% and deformation period of $T=200$), we observe that with increasing number of solvent particles, the biofilm is fully hydrated and the storage modulus  decreases (Fig. \ref{fig:Loss_Storage_Modulus}c). 
This is rather expected as under these experimental conditions the average distance between bacteria and polymers increases and the network becomes mechanically fragile (shear softening).
As clearly  observed from the data, the biofilm model is able to absorb the solvent. The more solvent is present, the softer  the storage modulus (Fig \ref{fig:Loss_Storage_Modulus}c). 
\noindent
On the other side, the hydration level seems not to  affect $G^{''}$, which only slightly increases.
This softening is not reflected in the loss modulus that remains more or less constant over the tested frequencies (Fig \ref{fig:Loss_Storage_Modulus}d). 
This implies that the present cross-linked matrix polymers might be entangeled, which allows expansion of the biofilm but at the same time provides rigidity.

\noindent
\subsection{Shear stress dependence on the shearing frequency: simulations versus experiments}
\begin{figure}[h!]
\centering
\includegraphics[scale=0.47]{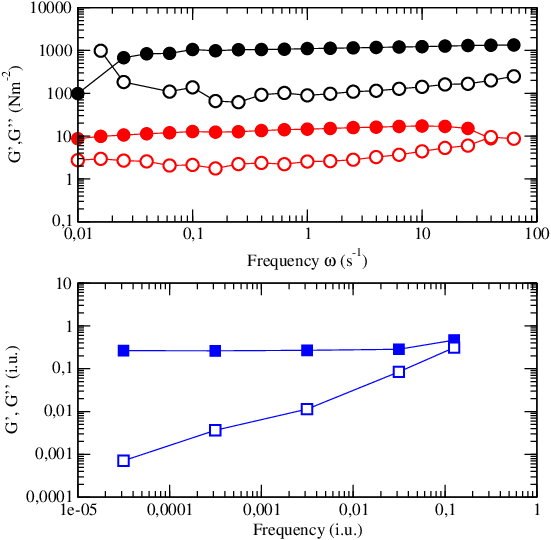}
\caption{Top) Experimental measurements of the elastic modulus (filled symbols) and loss modulus (empty symbols) for two different biofilms: one grown under shaking conditions (black) and one under static conditions (red). Bottom) $G^{'}$ (filled symbols) and $G^{''}$ (empty symbols) computed for the biofilm model. Data  refer to $N_b=184$, $N_p=80$, $l=100$, $N_s=35000$ and 2000 cross-links.\label{fig:fig_exp}
}
\end{figure}

\noindent
Finally,  we investigate the stress response as a function of  the frequency of the sinusoidal shearing strain $ \omega$. 
On the one side, we perform a rheology experiment on a  \textit{P. fluorescens} biofilm. 
We grow two \textit{P. fluorescens} biofilms for 24 hours: one under static conditions, and the second one under shaking conditions. 
Next, we locate the biofilm {\it ex vivo} in the rheometer and measure $G^{'}$ (filled symbols in Fig.\ref{fig:fig_exp}, top panel) 
and $G^{''}$ (empty symbols in Fig.\ref{fig:fig_exp}, top panel) 
for the static biofilm (Fig.\ref{fig:fig_exp}, top panel, red curves) and for the shaking biofilm  (Fig.\ref{fig:fig_exp}, top panel, black curves). On the other side, we perform a numerical rheology experiment on the model biofilm described earlier. 
After having studied the parameter space in  previous sections, we choose the following parameters: $CL=2000$, $N_b=184$, $N_p=80$, $l_p=100$, $N_s=35000$ and $A=75$ $\%$. 
We compute $G^{'}$ (filled symbols in Fig.\ref{fig:fig_exp}, bottom panel) and $G^{''}$ (empty symbols in Fig.\ref{fig:fig_exp}, bottom panel) as a function of the frequency of the externally imposed shear, and  
 compare the experimental wet-lab data with the biofilm simulations. \textit{P. fluorescens} biofilm grown after 24 hours under shaking conditions (in black) present higher shear moduli (both $G^{'} $ and $G^{''}$) than those measured for  \textit{P. fluorescens} biofilm grown under static conditions, independently on the frequency. Qualitatively, we observe this behavior for any \textit{P. fluorescens} biofilms grown longer than 24h (data not shown).


\noindent 
By increasing the externally imposed stress (or frequency), we observed increasingly larger absolute values of the $G^{''}$, which become comparable to $G^{'}$ for large frequencies (at least under static conditions, in red in Fig. \ref{fig:fig_exp}), as one would  expect in a  viscoelastic material.
In the biofilm grown under shaking conditions $G^{''}$ does not reach the same value of $G^{'}$ for large frequencies (in black in Fig. \ref{fig:fig_exp}). 
Due to the technical limitations of the used rheometer, we underline that  we could not measure the stress moduli of the \textit{P. fluorescens} biofilm grown under shaking conditions for frequencies beyond 100 Hz (Fig. \ref{fig:fig_exp}, top panel, black symbols).
However, we expect that eventually the trend of the stress response of the \textit{P. fluorescens} biofilms developed under shaken conditions will behave similarly to the observed stress response of the biofilm developed under static conditions (Fig. \ref{fig:fig_exp}, top panel).

 \noindent
The obtained values for the storage moduli (filled symbols) are always larger than the ones for the loss moduli (empty symbols), both in experiments and in simulations.
This result is consistent with our already published work\cite{jara2021}, where we showed that \textit{P. fluorescens} biofilms grow better  under shaking growth conditions than under static ones, the former leading to a 
higher bacterial density and a thicker biofilm matrix.
Augmented aeration and nutrient concentration within the biofilm provides an optimal growth conditions for the bacteria that eventually leads to the production of a larger amount of cross-links within the EPS polymer matrix and produces a stiffer biofilm structure \cite{jara2021}.

\noindent
According to the established experimental conditions, we numerically compute the shear moduli of  the biofilm model by changing the stress frequency over 4 order of magnitudes (Fig. \ref{fig:fig_exp}, bottom panel) . We assume that the main difference between the biofilm grown under static and shaking conditions is only in in the matrix composition, being the one of the shaken-grown biofilm richer in cross-links. 
In general terms, the numerical results (expressed in internal units, i.u.) exhibits the same overall trend (Fig. \ref{fig:fig_exp}, top panel)  observed for the \textit{P. fluorescens} biofilms (both static and shaken). 
Overall, for all tested frequencies,  $G^{'}$ is larger than $G^{''}$ that remains almost constant. However, $G^{''}$ increases with increasing frequency and eventually converges with $G^{'}$ at the highest simulated frequency. A qualitative difference is the drop in G'' shown by simulations at low frequencies. A dominance of the flowing solvent is the halmark for these simulations at low frequency of shearing. This is not however observed in experiments (see Fig. 4; top panel), which a change of the number of cross-links $CL$, bacteria $N_b$, polymers $N_p$ or hydration level $N_s$ within the studied parameter space  does not change the trend of our obtained computational results. A time-modulated solvent-matrix interaction could be introduced to correct this viscoelastic bias of simulations not found in experiments.

\section{Discussion}

Biofilms are composed of bacteria that secrete a network of extracellular polymeric substances (EPS) producing a viscoelastic matrix to which they eventually crosslink. Within a clinical or industrial environment, biofilm formation faces major issues. The mathematical modelling of complex biofilms might be of support,  providing  accurate descriptions and predictions that include physical parameters as hydrodynamics, solute mass transport or dynamics of the bacterial population within the biofilm. However, up-to-date  modelling of  biofilms, even at a coarse-grained level, has not been completely satisfactory. The parametrization of the individual components of a biofilm increases the complexity of the algorithm beyond computational capacity \cite{calude2017deluge}. Therefore, a simplification is needed that often leads to  covering  relevant experimental information. 

\noindent 
In our work, we present a mesoscopic coarse grained dissipative particle dynamics (DPD) approach to model  a mature bacterial biofilm. DPD models allow to simulate short range interactions of any particle and thus give a detailed picture of its close surroundings allowing an increase in spatial  resolution. This is an important advantage over generally used  long-range interaction models\cite{groot2004applications}.
Moreover the "soft" nature of DPD interactions allows to increase the integration time by few orders of magnitude, which permits to explore time scales  normally inaccessible to atomistic simulations. 
Most importantly,   our model  explicitly includes hydrodynamics and thermal fluctuations, which are known to play an important role in biological systems such as  biofilms \cite{krsmanovic2021hydrodynamics}. 

\noindent 
We generated a series of biofilm conformations by changing the composition,  the number of bacteria, the number of polymers or varied the cross-linking and the hydration level of the biofilm.
Next, the biofilm underwent  deformations  
when exposed to an external oscillatory shear with different amplitudes and frequencies spanning 4 order of magnitudes. We computed  the shear moduli and presented the model predictions for the different systems finding the computational results in agreement with experiments.
When varying the number of bacteria, we observed that $G^{'}$ remains constant throughout the $N_b$ range, in agreement with considering a biofilm as a polymer-colloid viscoelastic medium. 
The loss modulus $G^{''}$ increased linearly with $N_b$ reaching the values of the storage modulus for the largest number of bacteria. 
However, since $G^{''} $ is always smaller than $G^{'} $, the biofilm system behaves as a pasty material, with a relative high structural rigidity $G^{'} > G^{''} $.
When increasing the number of entangled polymers, we observed that it was easier for the biofilm to generate a solid-like network with an increased rigidity.  
The out-phase stress response  increases with increasing number of polymers.  However, since  the increase of $G^{''}$ is quite small when compared to the response caused by the presence of bacteria, we considered $G^{''}$ for polymer and solvent to be constant. 
Therefore, the present cross-linked matrix polymers might be entangeled, which allows expansion of the biofilm but at the same time provides rigidity.
When studying  the mechanical response of the system for different hydration levels, 
we observed that  increasing the number of solvent particles, the biofilm was fully hydrated and the storage modulus  decreased. The more the solvent, the softer  the storage modulus.


\noindent
We have chosen a set of parameter to compare our numerical data with rheological experiments performed on \textit{P. fluorescens} biofilms grown under static and shaken conditions. 
We choose to study biofilms formed by \textit{P. fluorescens} for two reasons.
1) It is the same model system we chose in our previous study\cite{jara2021}, and it will be easier for us to compare these results to the already published ones. 
2)Although the model microorganism used in this work for experimental data is considered non-pathogenic, other members of the genus \textit{P. fluorescens} such as \textit{Pseudomonas aeruginosa} is an opportunistic pathogen that is frequently associated with chronic biofilm infections. Overall, the composition of the EPS is similar in both species, with a high proportion of acetylated polysaccharides (alginate-like) and extracellular DNA \citep{Kives, Jennings}  and therefore suggested to expose a similar mechanical behaviour.
\noindent
Knowledge on the timing of mechanical transitions within biofilms may guide future strategies that allow the penetration of antimicrobial agents into “soften” biofilm matrices. 

\noindent
By accurately comparing the experimental to the numerical results, we conclude that 
our proposed coarse grained model catches qualitatively the moduli behaviors over four order of magnitudes 
as observed in Fig \ref{fig:fig_exp}. The measured elastic modulus is always higher in experiments and simulations than the loss modulus. Although $G^{'}$ and $G^{''}$ tend to converge for high frequency, the simulations contain more  viscous relaxation than reflected by experiments. Overall, the qualitative trends observed  as soft solids for both studied  biofilms (static and shaken) are recovered by our numerical findings, which could be eventually enhanced by including sticking details on the solvent-matrix interactions.
\noindent 
Moreover, this concordance exhibits the potential of the use of a coarse graianed DPD approach when  modelling  complex biofilms. Even though the model strongly relies on an {\it a priori} detailed study of the parameter space, in future prospective, this will allow us  to evaluate in more detail the biofilm transitions from a predominantly  solid-like system to predominantly liquid-like system, through the interplay of the elastic and loss moduli upon changing the biofilm composition.\\

\section*{Conflicts of interest}
There are no conflicts to declare.

\section*{Acknowledgments}

The authors acknowledge funding from Grant PID2019-105606RB-I00, 
FIS2016-78847-P, PID2019-108391RB-I00 (to FM), and FIS2015-70339 (to FM and ILM) ID2019-105343GB-I00 of the
MINECO and the UCM/Santander PR26/16-10B-2.
Francisco Alarc\'on acknowledges support from the ``Juan de la Cierva''
program (FJCI-2017-33580). AKM is recipient of a Sara Borrell fellowship (CD18/00206) financed by the Spanish Ministry of Health.
V.B. acknowledges the
support from the European Commission through the Marie
Skłodowska-Curie Fellowship No. 748170 ProFrost. 
The
authors acknowledge the computer resources from the Red Española de
Supercomputacion (RES) FI-2020-1-0015 and FI-2020-2-0032, and from the Vienna Scientific Cluster (VSC).

%


\end{document}